\documentclass[12pt,preprint,amsmath,amssymb,nofootinbib]{revtex4}

\usepackage{graphicx}
\usepackage{dcolumn}
\usepackage{bm}
\usepackage{subfigure}
\usepackage{cancel}
\newcommand{\be}{\begin{equation}}
\newcommand{\ee}{\end{equation}}
\newcommand{\ba}{\begin{eqnarray}}
\newcommand{\ea}{\end{eqnarray}}
\def\nonu{\nonumber}
\def\to{\rightarrow}

\begin{document}
\preprint{MADPH-09-1535}
\preprint{ZIMP-09-03}
\preprint{IPMU-09-0072}
\preprint{CAS-KITPC/ITP-131}

\title{ Like-sign Di-lepton Signals in Higgsless Models at the LHC}
\author{Tao Han~$^{\bf a}$}
\email{than@hep.wisc.edu}
\author{Hai-shan Liu~$^{\bf b}$}
\email{liu@zimp.zju.edu.cn}
\author{Ming-xing Luo~$^{\bf b}$}
\email{luo@zimp.zju.edu.cn}
\author{Kai Wang~$^{\bf c,d}$}
\email{kai.wang@ipmu.jp}
\author{Wei Wu~$^{\bf b}$}
\email{weiwu@zimp.zju.edu.cn}
\affiliation{
$^{\bf a}$ Department of Physics, University of Wisconsin, Madison, WI 53706, US
A\\
$^{\bf b}$ Zhejiang Institute of Modern Physics and Department of Physics, Zhejiang University, Hangzhou,
Zhejiang 310027, CHINA\\
$^{\bf c}$ Institute for the Physics and Mathematics of the Universe, University
 of Tokyo, Kashiwa, Chiba 277-8568, JAPAN\\
$^{\bf d}$ Kavli Institute for Theoretical Physics China, CAS, Beijing 100190, CHINA
}
\begin{abstract}
We study the potential LHC discovery of the $Z_{1}$ KK gauge boson unitarizing
$W^+_LW^-_L$ scattering amplitude. In particular, we explore the decay mode
$Z_{1}\to t\bar{t}$ along with $Z_{1}\to W^+ W^-$ without specifying the branching
fractions.
We propose to exploit the associated production $pp \to W Z_1$, and
select  the  final state of like-sign dileptons plus multijets and large missing energy.
We conclude that  it is possible to observe the $Z_1$ resonance
at a 5$\sigma$ level with an integrated luminosity of 100 fb$^{-1}$ at the LHC
upto  650 GeV for a dominant $WW$ channel, and 560 GeV for a dominant $t\bar t$ channel.
\end{abstract}
\maketitle

\section{Introduction}

Weak gauge boson masses are electroweak symmetry breaking (EWSB) effects and
how they arise still remains an open question in particle physics.
The minimal Higgs boson model provides a simple solution to the
electroweak symmetry breaking as well as fermion masses.
Without a scalar Higgs boson, the scattering amplitudes for the
longitudinally polarized gauge bosons ($W_L$ and $Z_L$) grow with
energy as $E^2$ and they violate the (perturbative) partial wave unitarity at
the energy scale $4\pi M_W/g\sim 1.5$~TeV \cite{unitarityhiggs}.
Unitarity bounds are also  reached at $16\pi/\sqrt{2} G_F M_f \xi$
for massive fermion scattering amplitudes $f\bar{f} \to W^+_L W^-_L$ \cite{ac},
where $\xi=1$ for leptons and $\xi=\sqrt{3}$ for quarks. Due to the large
top quark mass, the scale of mass generation for the top quark yields the strongest bound,
to be about 3.5 TeV \cite{topunitarity}.

Recently there have been attempts to generate EWSB without Higgs bosons in
the framework of extra dimension models \cite{higgsless}. In the so-called
``Higgsless"  models, unitarity is restored by contributions from a
tower of Kaluza-Klein (KK) gauge bosons.
The $E^2$ dependence in the scattering amplitude of $W_L$ and $Z_L$ is
cancelled by the Kaluza-Klein (KK) modes $Z_n$, $W_n$.
The unitarity in massive fermion scattering will also be protected by
allowing $Z_n t\bar{t}$.

There have been several studies regarding
the LHC search of the first KK $W_n$ boson ($W_1$) via weak boson
fusion (WBF) \cite{Birkedal:2004au,He:2007ge} and associated
production \cite{He:2007ge}. These studies all focused on the
decay channel $W_1\to WZ$ and chose multilepton final states.
Despite the small $Z$ leptonic decaying branching
fraction (BR), the SM backgrounds are always under control.
In WBF $jjW_1\to jjW^\pm Z$
through the $3\ell+jj+\cancel{E}_T$ final state, the claimed reach
is 1 TeV for 5$\sigma$ discovery at 100 fb$^{-1}$ integrated
luminosity \cite{He:2007ge}. For the associated
production $ZW_1\to W^\pm ZZ \to 4\ell+jj$ final state had been
proposed and the claimed reach is about 620 GeV at  5$\sigma$ for 100
fb$^{-1}$ integrated luminosity \cite{He:2007ge}.

In this paper,  we will explore another aspect of this class of model.
Given the large top quark mass, we argue that
$Z_1$ should couple to $t\bar{t}$ significantly \cite{tophiggsless}
if $Z_1$ is the dominant contribution  of unitarizing the scattering amplitudes involving top pairs.
We take a phenomenological approach to the $Z_{1}$ boson, and allow it to decay to
both $W^+ W^-$ and $ t\bar{t}$ channels
without specifying the decaying BRs. Focusing on the $W^+W^-Z_1$
coupling, the $Z_1$ can be produced through WBF channel $jj Z_1$ and
associated production $WZ_1$\footnote{Associated production of a strongly interacting vector
($V$) with heavy quarks has been considered in Ref.~\cite{Han:2004zh}. The perspectives for its observation
at the LHC look promising.}.
Allowing $Z_{1}\to W^+ W^-$/$Z_{1}\to
t\bar{t}$ decays, the final states will contain
$jjt\bar{t}$/$jjW^+W^-$ and $W^\pm t\bar{t}$/$W^\pm W^+W^-$. Unlike
the $W_1$ case in \cite{Birkedal:2004au,He:2007ge}, the WBF channel $jjZ_1$
suffers from huge irreducible Standard Model (SM)
background of $t\bar{t}$ plus jets. Our studies focus on the leptonic decays of like-sign $W$s
via associated production.
Consequently, the final states to be studied consist of multijet $+\ell^\pm
\ell^\pm+\cancel{E}_T$. In comparison with previous studies,
we can reach a mass of 550 GeV$-$650 GeV for $t\bar t$ and $WW$ modes,
at a 5$\sigma$ level with 100 fb$^{-1}$
of the integrated luminosity.

This paper is organized as follows. In Sec.II, we discuss the
parameters in the model, bounds on KK gauge boson
masses and their couplings to SM particles due to
unitarity requirement in scattering amplitudes and
precision electroweak tests. In Sec.III,
we discuss the search of $Z_1$ gauge boson at the LHC
allowing both $Z_1\to W^+W^-$ and $Z_1\to t\bar{t}$ decays
and focus on the multijet $+\ell^\pm \ell^\pm+\cancel{E}_T$ final
states. We will conclude in Sec. IV.

\section{ Model Parameters}

In Higgsless models, it is shown \cite{unitarityextradim} that
the scattering amplitude of the longitudinal components of
massive $W$ and $Z$ bosons can be unitarized by
the KK excitations of the gauge bosons $Z_{n}$ and $W_{n}$.
The cancellation of  $E^2$ and $E^4$ terms lead to the following
sum rules of $Z_{n}$ as \cite{higgsless}
\ba
  g_{WWWW}^2&=&g_{WWZ}^2+g_{WW\gamma}^2+\displaystyle \sum_n(g_{WWZ_{n}})^2~, \label{eq:zpsum1}\\
  4g_{WWWW}^2M_W^2&=&3\Big[g_{WWZ}^2 M_Z^2 + \displaystyle \sum_n(g_{WWZ_{n}})^2 M_{Z_{n}}^2\Big]~, \label{eq:zpsum2}
\ea where $g_{WWWW}=g^2=e^2/\sin{\theta_W}$ is the SM four $W$
contact interaction coupling,  $g_{WWZ}$/$g_{WW\gamma}$ are the SM
coupling between $WW$ and $Z$/$\gamma$ respectively.

We focus on the first KK excitation $Z_1$ boson and assume
other higher KK excitations to be less relevant to the collider search.
However, such a truncation will lead to a violation of partial wave unitarity
at a scale $\Lambda \simeq 4\pi M_{Z_{1}}/g$ \cite{higgsless}.
For instance, if $M_{Z_{1}} = 500$ GeV, then $\Lambda$ is of ${\mathcal O}(10$ TeV),
which should not affect our phenomenological considerations.

In principle, the introduction of new gauge boson will also modify the
SM couplings. We keep $g_{WW\gamma}$ and $g_{WWWW}$ to their SM value
and $g_{WWZ_1}$ and $g_{WWZ}$ can be computed
through Eq. \ref{eq:zpsum1} and Eq. \ref{eq:zpsum2}. We obtain
the couplings as
\ba
  g_{WWZ_1}^2 &=& \frac{(4M_W^2-3M_Z^2)g_{WWWW}^2+3M_Z^2 g_{WW\gamma}^2}{3 (M_{Z_{1}}^2- M_Z^2)}~, \label{eq:gwwzp}\\
  g_{WWZ}^2 &=& \frac{3M_{Z_{1}}^2 (g_{WWWW}^2-g_{WW\gamma}^2)-4M_W^2 g_{WWWW}^2}{3(M_{Z_{1}}^2-M_Z^2)}~.\label{eq:gwwz}
  \ea
The $Z_{1}$ mass $M_{Z_{1}}$ will then be the only input parameter in the analysis.
If $M_{Z_{1}} = 500~\text{GeV}$, the deviation of $g_{WWZ}^2$ from its SM value is smaller than $1\%$,
which is consistent with the current experimental bound.

Previous studies have shown that it is hard to accomodate precision
electroweak data if $Z_1$ couples to the SM fermions. Even if
$Z_1$ only couples to the third family, it is disfavored by the
strong constraints on the $Zb\bar{b}$ coupling. A fermiophobic
$Z_1$/$W_1$ have been studied in
\cite{SekharChivukula:2006cg,Chivukula:2005xm}.
A few viable models have been
suggested to incorperate this \cite{tophiggsless,terning}.
In this paper, we will take $Z_1$ to couple mainly
to $t\bar{t}$ and $W^+W^-$ but
without specifying the values of the branching fractions
$\text{BR}(Z_1\to t\bar{t})$ and $\text{BR}(Z_1\to W^+W^-)$ a priori.

\section{$Z_{1}$ at the LHC}

\begin{figure}[tb!]
    \includegraphics[scale=1.0,width=9cm]{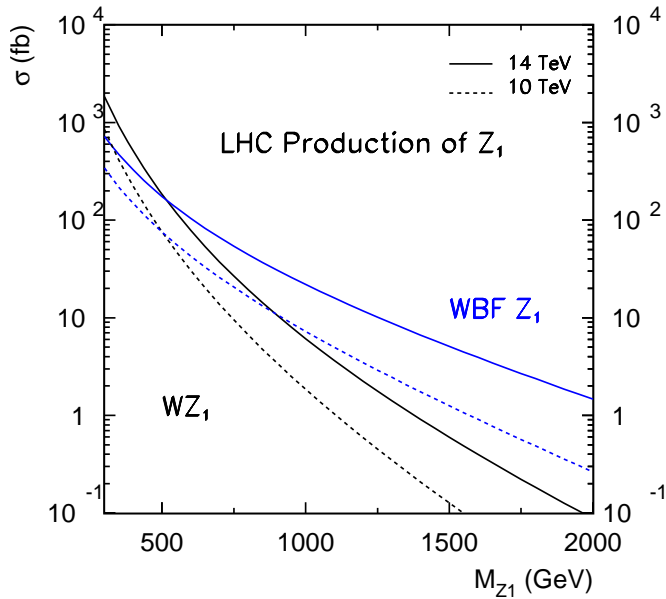}
    \caption{\label{fig:rate} $Z_{1}$ Production rates
    at the LHC for 10 (dotted curves) and 14 TeV (solid curves).  }
\end{figure}

For a vector state, the dominant production mechanism is the Drell-Yan process, $q\bar q\to Z_1$,
if there are sizable couplings to light quarks \cite{foursite}.
With highly suppressed couplings to light fermions as preferred in Higgsless models,
the dominant production channel
for $Z_1$ is commonly considered as the weak boson fusion (WBF)
mechanism  \cite{Birkedal:2004au,He:2007ge,Belyaev:2009ve}
$p p \rightarrow Z_{1} j j$. We extend the calculation by including the
the associated production $p p \rightarrow W Z_{1}$.
Fig.~\ref{fig:rate} shows the total cross sections of $Z_{1}$ from
associated production and WBF production at the LHC for 10 (dotted
curves) and 14 TeV (solid curves). We have chosen the PDF  CTEQ6L
and the factorization scale $\mu_F = \sqrt{\hat{s}}/2$. As expected,
we see that the associated production is larger at smaller value of
$M_{Z_1}$ and the WBF mechanism takes over at higher values. The two
curves cross near $M_{Z_1}\sim 400-600$ GeV, where the cross
sections are quite sizable,  about 100$-$200 fb.

%
The WBF processes for $W_LW_L$ scattering have unique kinematic
features: two forward/backward energetic jets with large dijet
invariant mass and lack of central jet activities.
However, the $Z_{1}\to t\bar{t}$ and $Z_{1}\to W^+W^-$ channels via WBF
suffer from huge backgrounds, mainly from $t\bar{t}$ plus jets of order several hundreds pb,
while the $Z_1$ WBF production rate is only of
a few hundreds fb which is ${\mathcal O}(10^3)$ smaller.  The $W Z_{1}$
associated production, however, benefits from the additional handle of $W^\pm$.
We therefore focus on this associated production channel.
The processes under consideration are
\be
pp\to W^\pm Z_{1} \to W^\pm\  t\bar{t}\ \ \text{and}\ \  W^\pm\  W^+ W^-.
\ee
As for the final state reconstructions, although the hadronic decay of $t$ or $W$ will help to
fully reconstruct the $Z_{1}$ resonance, these channels suffer from
huge standard model background from $t\bar{t}$ plus jets. To effectively suppress the backgrounds,
we look for the signal of like-sign di-leptons.
We choose the $W^\pm$ associated with the $Z_{1}$ production
always decaying into leptons ($\ell=e,\ \mu$)
and take the $W^\pm$ from $Z_1$ decay of the like-sign as the previous $W^\pm$, which also decays leptonically.
The third $W^\mp$ decays hadronically.
The final states that we are looking for are
\ba
pp\to W^\pm Z_{1}  \to  \left\{
\begin{array}{ll}
W^\pm\  W^+ W^-\to \ell ^\pm \ell ^\pm +jj+\cancel{E}_T , \\
W^\pm\  t\bar{t}\to \ell^\pm \ell^\pm+ b\bar{b} \ jj +\cancel{E}_T,
\end{array}
\right.
\ea
with $jj$ always reconstruct on-shell $W$. The BR then carries a factor of
\be
\left({2\over 9}\right)^2\times {2\over 3} = 3.29\% ,
\ee
to be multiplied by $\text{BR}(Z_{1}\to t\bar{t})$ or $\text{BR}(Z_{1}\to W^+W^-)$ in addition.

To simulate detector effects on energy-momentum
measurements, we smear the electromagnetic energy and lepton
momenta by a Gaussian distribution whose width is parameterized as
 \cite{CMS}
\begin{eqnarray}
{ \Delta E\over E} &=& {a_{cal} \over \sqrt{E/{\rm GeV}} } \oplus b_{cal}, \quad
a_{cal}=5\%,\quad b_{cal}=0.55\% .
\label{ecal}
\end{eqnarray}
We did not separately smear the muon $p_T$ by tracking resolution,
since separate smearing do not affected the results practically.
The jet energies are also smeared using the same Gaussian formula as in
Eq.~(\ref{ecal}), but with  \cite{CMS}
\begin{equation}
a_{cal}=100\%,\quad  b_{cal}=5\%.
\end{equation}

\subsection{$Z_1 \to WW$}

\begin{figure}[t]
    \includegraphics[scale=1.0,width=8cm]{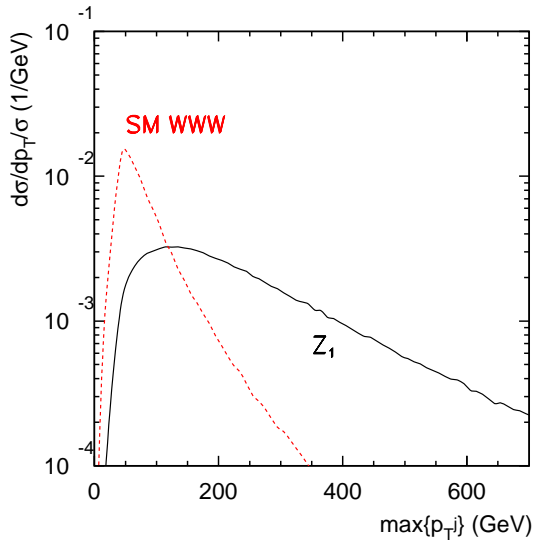}
    \caption{\label{fig:ptjmax} Normalized $\text{max}(p^j_T)$ distribution for $2j+\ell^\pm \ell^\pm +\cancel{E}_T$
    final states of $W^\pm Z_{1}\to W^\pm W^+W^-$ and the SM $W^\pm W^+W^-$ at $M_{Z_{1}}= 500$ GeV after $M_{jj}$ cut.}
\end{figure}

We first consider the $Z_{1}\to W^+W^-$ channel for the like-sign dilepton
plus dijet $\ell^\pm \ell^\pm +jj+\cancel{E}_T$ final state.
We propose the basic cuts for the event selection as:
  \ba
  p^j_T > 25 ~\text{GeV} &;& |\eta_j|<3.0 \nonu\\
  p^\ell_T > 15 ~\text{GeV}  &;& |\eta_\ell|< 2.5
  \label{eq:basic} \\
  \Delta R(j,j)> 0.4 &;&\Delta R(j,\ell)>0.4. \nonu
  \ea
We further demand that the dijet in our signal reconstruct an on-shell $W$ boson
 \be
  |M_{jj}-M_W| < 15~ \text{GeV}.
  \label{eq:mjj}
 \ee
After these selection cuts, the leading QCD background ($uu,\ \bar
d\bar d \to W^\pm W^\pm\ jj$) is  reduced to a negligible level. The
remaining background is mostly the electroweak $WWW$ production with
two like-sign $W$ decaying leptonically and the third $W$ decaying hadronically.
At this stage, this $WWW$ background rate is already
smaller than that of the signal if $\text{BR}(Z_{1}\to W^+W^-)\sim
100\%$.
We note that the decay products from a heavy resonance $Z_{1}$ will
be fairly energetic, with a typical transverse momentum $p_T^j\sim
0.5 M_{Z_1} \sqrt{1-4M_W^2/M_{Z_1}^2 }$. This is shown in
Fig.~\ref{fig:ptjmax}, where we plot the hardest jet $p_T$. We can
further improve the signal purity by imposing a cut , for instance,
\begin{equation}
{\rm max}(p^j_T) > 150\ {\rm GeV}.
\label{eq:ptj}
\end{equation}
We show the results for the signal with $M_{Z_{1}}=500~\text{GeV}$
and $\text{BR}(Z_{1}\to W^+W^-) = 100\%$ and backgrounds in Table
\ref{table:result2}. With a given number of events for the signal
($S$) and background $(B)$, we conservatively estimate the statistical significance
by
\begin{equation}
S/\sqrt{S+B}.
\label{eq:SB}
\end{equation}
We see that the signal observability is quite convincing with 10$-$30 fb$^{-1}$ even before the final cut
of Eq.~(\ref{eq:ptj}). However, with this additional cut, the $S/B$ is significantly improved from 2 to 8,
making the systematics of the measurement much  less a concern.
If we take the BR($Z_1$) as a free parameter, we can see that requiring a 5$\sigma$ signal sensitivity, one is
able to probe the BR($Z_1\to WW$) to a level of $54\%$ with 30 fb$^{-1}$.

\begin{table}[t]
  \begin{tabular}{|c| c| c| c| c|}
  \hline
  \ &No Cut&Basic Cut Eq.~(\ref{eq:basic}) & $+M_{jj}$ Eq.~(\ref{eq:mjj}) & +max$(p_T^j)$ Eq.~(\ref{eq:ptj})\\
  \hline
  Signal (fb)& 6 &2.8 &2.7 & 1.7\\
  \hline
$W^\pm W^\pm jj$ BG (fb)& 41 &17 &0.33 & 0.03\\
  \hline
$W^\pm W^\pm W^\mp$ BG (fb)& 4.1 & 1.1 & 1.1 & 0.18 \\
  \hline
Total BG (fb) & 45 & 18 & 1.5 & 0.21 \\
  \hline
\hline
  S/B&\ & &1.9 & 8.1 \\
  \hline
  S/$\sqrt{S+B}$ at 10 fb$^{-1}$ &\ & & 4.2 & 3.9 \\
    S/$\sqrt{S+B}$ at 30 fb$^{-1}$ &\ & & 7.2 & 6.8 \\
\hline
  \end{tabular}
  \caption{Signal/Background Comparison in $p p \to \ell^\pm\ell^\pm+2j+\cancel{E}_T$ final states
  $M_{Z_{1}}= 500$ GeV and $\text{BR}(Z_{1}\to W^+W^-) = 100\%$.  }
    \label{table:result2}
\end{table}

%

\begin{figure}
    \includegraphics[scale=1.0,width=8cm]{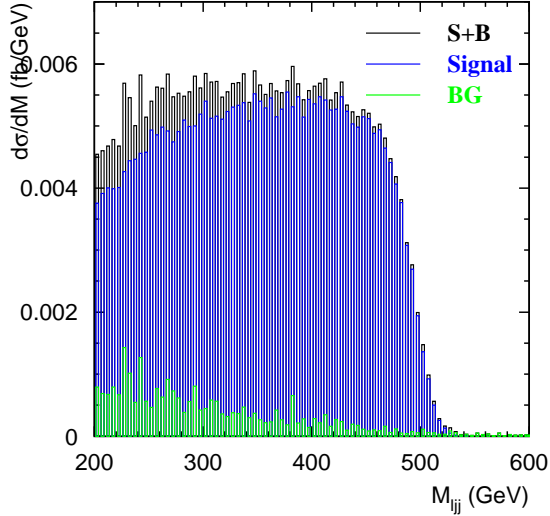}
    \caption{\label{fig:m3} $M_{\ell jj}$ distribution for signal at $M_{Z_1}=500$~GeV and background after $M_{jj}$ and $\text{max}(p^j_T)$ cuts. }
\end{figure}

There are two missing neutrinos in the final states that we propose
and only one of them is from the resonance. Consequently, the
reconstruction of the resonance $Z_{1}$ is very challenging. The
like-sign dilepton final states will also cause combinatorial
problem as it is difficult to distinguish a lepton from the $W^\pm$
in $W^\pm Z_{1}$ or from the $Z_{1}$.
%
Between the two reconstructed invariant masses $M_{\ell jj}$,  we
propose to use the smaller one since this will be bounded by the
resonance mass $M_{Z_{1}}$. Figure~\ref{fig:m3} shows the distribution
of the smaller $M_{\ell jj}$ for $M_{Z_{1}} = 500$ GeV after $M_W$
reconstruction and $\text{max}(p^j_T)$ cut. We see a clear endpoint
near $M_{Z_1}$. Because of the limited statistics, we would not
impose further cut on this variable, although it could help to
determine the mass of $Z_1$.

\subsection{$Z_1 \to t\bar t$}

\begin{figure}[t]
    \includegraphics[scale=1.0,width=8cm]{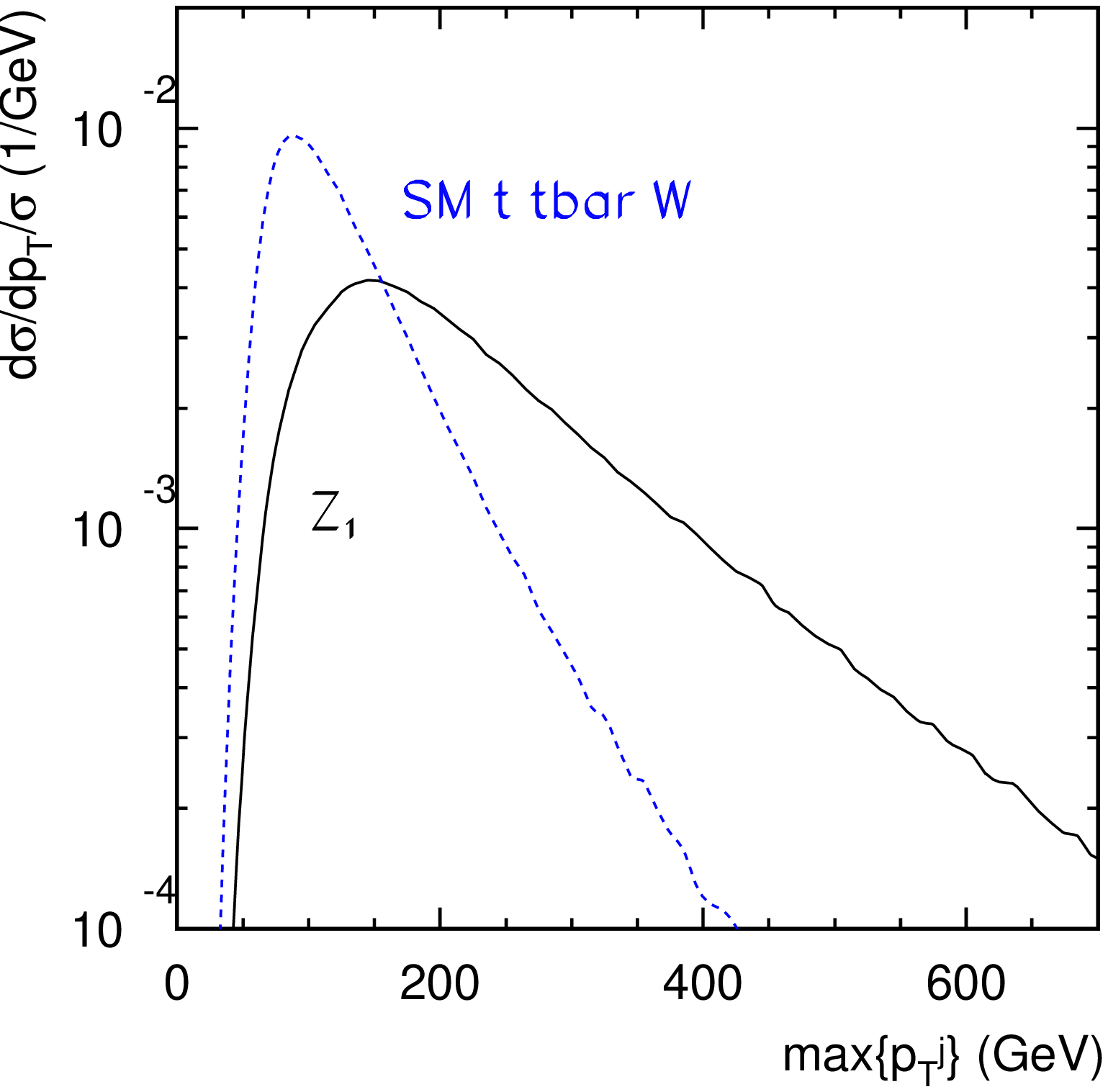}
    \caption{\label{fig:ppwzpptjmax} Normalized $\text{max}(p^j_T)$ distribution of  $p p\to
W^\pm Z_1 \rightarrow t \bar{t} W^\pm \to 4j +\ell^\pm \ell^\pm +\cancel{E}_T$ at $M_{Z_1}=500$~GeV and SM $t\bar{t}W^\pm$ background. }
\end{figure}

\begin{table}[t]
  \begin{tabular}{|c| c| c| c| c|}
  \hline
  \ &No Cut&Basic Cuts& $+\text{max}(p^j_T) > 150\ \text{GeV} $ & +$M_{\ell 4j}$ cut\\
  \hline
  Signal(fb)& 6 &1.4 & 0.91 & 0.64 \\
  \hline
  SM BG(fb)&14 & 3.6 &1.1& 0.25\\
  \hline
    \hline
  S/B&\ &0.40& 0.84& 2.5\\
  \hline
  S/$\sqrt{S+B}$ at 30 fb$^{-1}$&\ & 3.4 &  3.5 & 3.7 \\
    S/$\sqrt{S+B}$ at 100 fb$^{-1}$&\ &6.2 & 6.4& 6.8\\
  \hline
  \end{tabular}
  \caption{Summary Table of $pp\to W^\pm Z_{1}\to W^\pm t \bar{t}\to  4j +\ell^\pm \ell^\pm +\cancel{E}_T$ for $M_{Z_{1}}=500$ GeV and its leading SM background $t\bar{t}W^\pm$.}
  \label{table:result1}
\end{table}

In the case where $Z_{1}\to t\bar{t}$ has significant decay branching fraction,
we will look for $\ell^\pm \ell^\pm+bbjj+\cancel{E}_T$ final states. We adopt the same event
selection as in Eq.~(\ref{eq:basic}).
The leading irreducible SM background comes from the process $t\bar{t}W^\pm$
which also contains two $b$-jets. Therefore, a requirement of $b$-tagging in our study
would not help to reduce this background.
The first handle we exploit is again the boost effects from heavy
resonance $Z_{1}$. Fig.~\ref{fig:ppwzpptjmax} shows the
$\text{max}(p^j_T)$ distributions for SM $t\bar{t}W^\pm$ and $W^\pm
Z_{1} \to t\bar{t} W^\pm$.
We thus impose the same cut as in Eq.~(\ref{eq:ptj}) on max($p_T^j$).

\begin{figure}
    \includegraphics[scale=1.0,width=8cm]{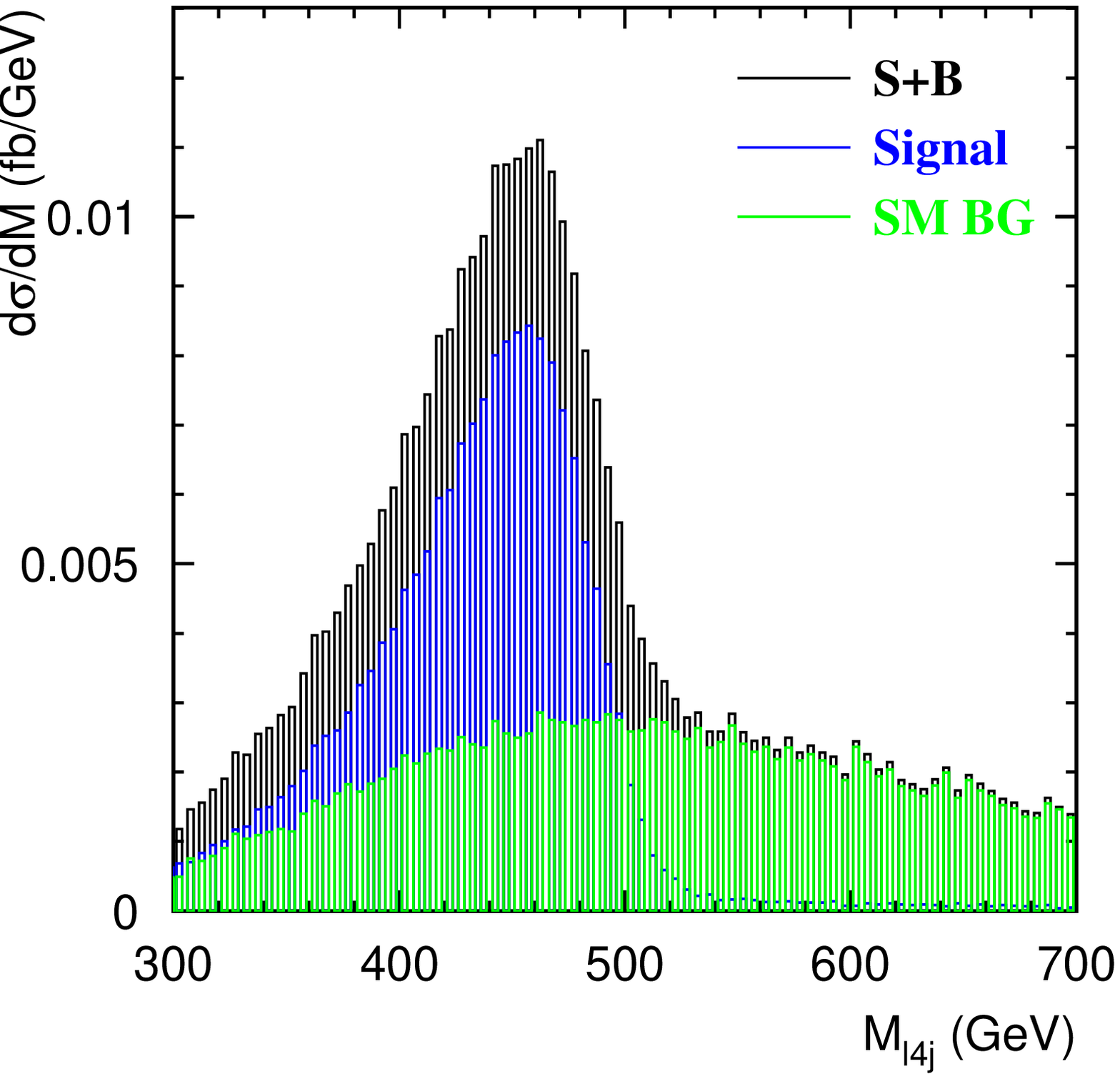}
    \caption{\label{fig:m5} $M_{1l4j}$ distribution of $p p\to
W^\pm Z_1 \rightarrow t \bar{t} W^\pm \to 4j +\ell^\pm \ell^\pm +\cancel{E}_T$ at $M_{Z_1}=500$~GeV and SM $t\bar{t}W^\pm$ background. }
\end{figure}

As for the $Z_1$ mass reconstruction from $t\bar t$,
we encounter the same problem of the combinatorics
as before. The two on-shell top quarks provide extra handle in reconstruction.
One can first try to reconstruct the hadronic top quark then the $Z_1$ resonance.
We first require one of the reconstructed $M_{jjj}$'s to be close to top quark mass to reconstruct
the hadronic decaying top. The other jet will be combined with
the two leptons to make $M_{\ell j}$ invariant mass.
Similar to the $Z_{1}\to WW$ search, we propose to explore the invariant mass
distribution for one of the leptons plus the fourth jet, and choose the smaller
invariant mass between the two $M_{\ell j}$ to reconstruct the leptonic decaying
top quark. We wish to consider the reconstructed top pair invariant mass $M_{tt}$ with the
endpoint related to $M_{Z_{1}}$.
We plot the smaller $M_{\ell 4j}$ distribution in Fig.~\ref{fig:m5},
after the $\text{max}(p^j_T)$ cut. This motivates us to select the mass window to estimate
the accessible sensitivity to the signal, and we choose
\begin{equation}
0.8 M_{Z_{1}} <M_{\ell 4j} < M_{Z_{1}}.
\end{equation}
The results are summarized in Table \ref{table:result1} with variety of cuts.
We see that the statistical significance is also convincing with 50$-$100 fb$^{-1}$.

\begin{figure}
    \includegraphics[scale=1.0,width=7.5cm]{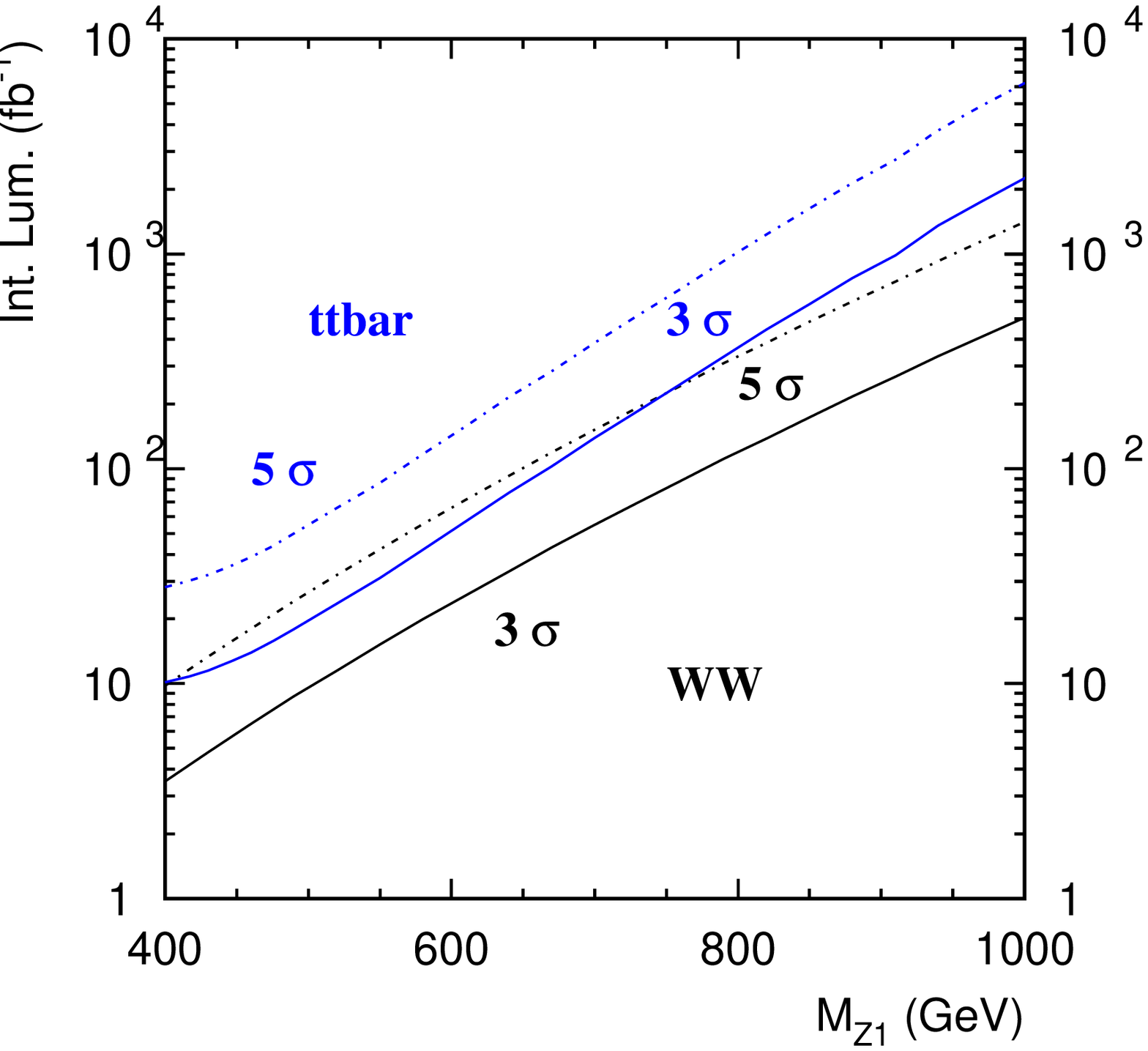}
        \includegraphics[scale=1.0,width=7.5cm]{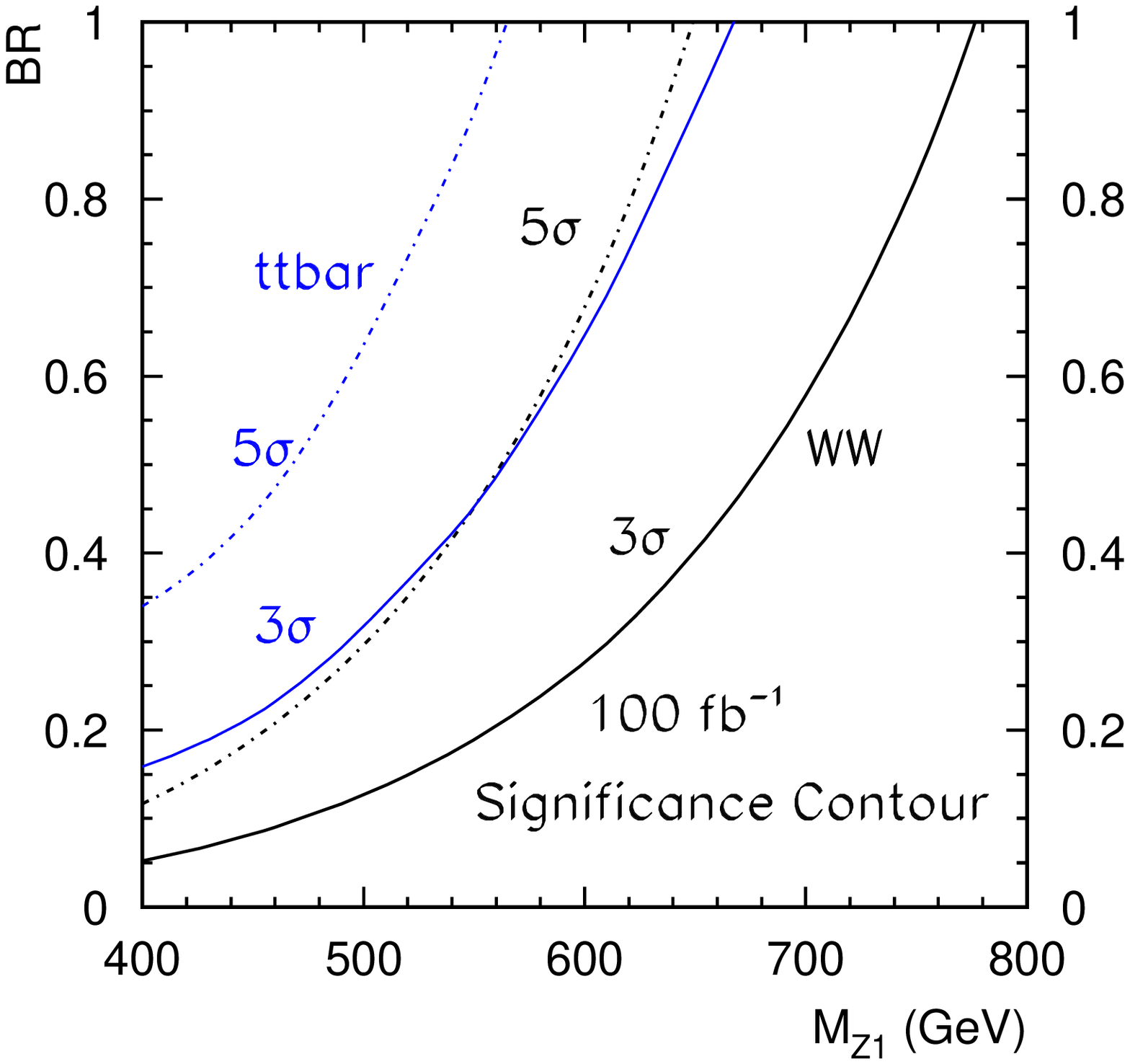}
    \caption{\label{fig:detection3} (a) Intergrated luminosities required for $3\sigma$ and $5\sigma$ significance
    of detection after $Z_{1}$ reconstruction; (b) Accessibility to BR($Z_1\to WW$)  and  BR($Z_1\to t\bar t$)
    at  the $5\sigma$ level  with 100 fb$^{-1}$. }
    \end{figure}

To summarize this section, in Fig.~\ref{fig:detection3}(a) we plot
the integrated luminosity needed to reach a 3$\sigma$ (solid lines)
and 5$\sigma$ (dashed lines) statistical significance versus
$M_{Z_{1}}$, with a reconstruction window $0.6 M_{Z_{1}} <M_{\ell
2j} < M_{Z_{1}}$, which helps more for a heavier $Z_1$. To claim a
3$\sigma$ discovery of the $Z_1\to W^+W^-$ for $M_{Z_1}= 1$ TeV, it would
require 500 fb$^{-1}$ integrated luminosity.  We also plot the BR
parameter reached at 3$\sigma$ (solid lines) and 5$\sigma$ (dashed
lines) level with  a 100 fb$^{-1}$ integrated luminosity in
Fig.~\ref{fig:detection3}(b), again  versus $M_{Z_{1}}$. One can
reach 60\% BR with 3$\sigma$ significance  for 700 GeV and 580 GeV of $Z_1$ decay to
$WW$ and $t\bar t$ channels, respectively. We adopt the criterion of Eq.~(\ref{eq:SB}) for estimations.

If the $Z_{1}$ is heavier, for the $Z_{1}\to W^+W^-$ channel,
the hadronic decaying $W$ from $Z_{1}$ will
be highly boosted and one may define one fat $W$-jet, of which the jet mass
is within $M_W$ window. The signature will then become
$\ell^\pm \ell^\pm+ J_W+\cancel{E}_T$. Although it would be challenging to quantify
this background without detailed simulation of the detector effects, one may expect that the
background should be smaller  as studied in \cite{type2}, and in particular there will be no
leading process $W^\pm W^\pm j$ in the SM.
If $Z_{1}\to t\bar{t}$ is dominant, the coverage is only up to 650 GeV or so for 100 fb$^{-1}$
integrated luminosity. The top quarks decaying from the $Z_{1}$ are not
boosted into one top-jet cone. For larger $Z_1$ mass, if top decay becomes highly
boosted, the discovery will become more challenging as we won't have the isolated
like-sign dilepton any more.

\section{Conclusion}

We have studied the LHC phenomenology of $Z_{1}$ that unitarize the $WW$
scattering amplitude. $Z_{1}$ does  not couple to light fermions
but we allow it to couple to the top quark.  We choose the associated
production $W^\pm Z_{1}$ and study both $Z_{1}\to W^+ W^-$ and
$Z_{1}\to t\bar{t}$ without specifying the decaying BRs. By choosing
the multijet + $\ell^\pm \ell^\pm$+ $\cancel{E}_T$ final state, we
find that it is quite feasible for the signal to be larger than the SM
irreducible background. Even though it is hard to fully reconstruct
the resonance $Z_{1}$, we propose to use the edge of jets plus lepton
invariant mass $M_{\ell nj}$ to get some information of the
resonance $Z_{1}$ mass. For 100 fb$^{-1}$ integrated luminosity,
assuming 100\% decay BR, for 5$\sigma$ discovery significance, the
$Z_{1}\to t\bar{t}$ can be searched upto 560 GeV and  $Z_{1}\to W^+
W^-$ search can reach 650 GeV respectively.

\section*{Acknowledgement}
T.H. is supported in part by the U.S. DOE under Grants
No.DE-FG02-95ER40896, W-31-109-Eng-38, and in part by the Wisconsin
Alumni Research Foundation. HL, ML and WW are supported in part by
the National Science Foundation of China (10425525) and (10875103).
KW is supported by the World Premier International Research Center
Initiative (WPI Initiative), MEXT, Japan and in part by the Project
of Knowledge Innovation Program (PKIP) of the Chinese Academy of
Sciences.

\end{document}